\documentclass[conference]{IEEEtran}
\ifCLASSINFOpdf
 \usepackage[pdftex]{graphicx}
\else
\fi

\usepackage{amsmath}
\usepackage{times}
\usepackage{pifont}
\usepackage{txfonts}

\begin{document}
%
\title{Measuring the performance of sensors that report uncertainty}

\author{\IEEEauthorblockN{A. D. Martin and T. C. A. Molteno}
\IEEEauthorblockA{
Department of Physics\\ 
University of Otago\\ 
Dunedin, 9016, New Zealand\\
Email: amartin@elec.ac.nz
}
\and
\IEEEauthorblockN{M. Parry}
\IEEEauthorblockA{
Department of Mathematics and Statistics\\ 
University of Otago\\ 
Dunedin, 9054, New Zealand}}


\maketitle

\begin{abstract}
We provide methods to validate and compare sensor outputs, or inference algorithms applied to sensor data, by adapting statistical scoring rules. The reported output should either be in the form of a prediction interval or of a parameter estimate with corresponding uncertainty. Using knowledge of the `true' parameter values, scoring rules provide a method of ranking different sensors or algorithms for accuracy and precision. As an example, we apply the scoring rules to the inferred masses of cattle from ground force data and draw conclusions on which rules are most meaningful and in which way.
\end{abstract}


%
\IEEEpeerreviewmaketitle

\section{Introduction}\label{Sec:Intro}

In principle, no measurement of a continuous physical quantity can be made to absolute precision. In accordance with this fact many sensors will quote a precision or resolution; more sophisticated sensors may return an estimate of uncertainty for each individual measurement. A near-optimal measurement from such a sensor will comprise a parameter estimate close to the `true' value with uncertainty overlapping this `true' value while remaining as small as possible. 
Sensor uncertainty quantification has been discussed in the literature, e.g., in optical time-of-flight sensors \cite{Edeler:2014}, nuclear power plant monitoring
 \cite{Ramuhalli:2013}, magnetometers \cite{Bernieri:2007} and databases of drifting sensors \cite{Cheng:2003}; and methods of dealing with such uncertainty have been discussed in, e.g.,  Refs.\ \cite{Frew:2004, Bis:2009}. However, an extensive search of the literature has revealed no work on validation of sensor uncertainty.

In this paper, we provide methods  validating and comparing sensor performance given knowledge of the `true' parameter value,  which capture the properties of optimal measurements. Testing either the measurement accuracy or the uncertainty prediction individually is relatively straightforward: one could test for accuracy over many runs by finding, e.g., the rms distance between the prediction and the `true' value; however, this tells us nothing about quality of the uncertainty reporting. Likewise, the uncertainty reporting could itself be validated over many runs, e.g., by measuring the proportion of the prediction intervals which contain the `true' value, but this tells nothing about the accuracy of the measurement. In order to produce single measures which reward both accuracy and precision, we turn to statistical scoring rules \cite{Rafferty:2007}. Scoring rules are used to score probabilistic forecasting methods  in meteorology \cite{Jolliffe:2003, Brocker:2007}, climate models \cite{Suckling:2013}, economics \cite{Boero:2011, Gneiting:2011}, and football predictions \cite{Constantinou:2012}. 

A scoring rule is a function $S(q|x)$ which returns a score for distribution $q$, given the `true' value $x$. A proper scoring rule is one whose expectation value $\int S(q|y)p(y) dy$ is optimised  when $q=p$. (In this paper we use the convention that the best estimate gains the minimum score.) Such proper scoring rules encourage users to supply their best estimate of the distribution, since they cannot be `gamed' by supplying an alternative  \cite{Rafferty:2007}. A scoring rule is said to be strictly proper when $p$ uniquely minimizes the expectation value. In this paper we adapt scoring rules to operate on typical sensor outputs, and analyse their characteristics.

This paper has the following format: in Sec.\ \ref{Sec:SensorOutput} we discuss the form of typical sensor output; in Sec.\ \ref{Sec:ScoringRules} we adapt scoring rules for such outputs and discuss their properties; in Secs.\ \ref{Sec:Experiment} and \ref{Sec:Results} we discuss an experiment we use as a case study to demonstrate the scoring rules and in Sec.\ \ref{Sec:Conclusions} we draw conclusions about the uses and relative qualities of different scoring rules.

\section{Sensor output}\label{Sec:SensorOutput}
We adapt scoring rules for validating sensor (or algorithm) output which incorporates an estimate of uncertainty in the measurement of a physical parameter.  Such an estimate could be in the form of a Bayesian posterior distribution for that parameter, or some other estimate of measurement accuracy. For example, a parameter estimate from the output of a Kalman filter \cite{Kalman:1960} will consist of the mean and variance characterising the posterior Gaussian distribution. For practical reasons, rather than reporting a complete probability distribution, we would expect a sensor to report only a prediction interval or a prediction and error prediction.
Therefore we formulate the scoring rules to act on only either the mean and variance of a predicted distribution or an interval $\left[ x_1, x_2\right]$, in which the `true' value is estimated to reside (ideally with a specified probability). Such an interval could represent, for example, the region between the 5th and 95th percentiles of a posterior distribution.

\section{Scoring Rules}\label{Sec:ScoringRules}
The scoring rules require knowledge of the `true' parameter value, $x$, obtained, for example, by a method with negligibly small uncertainty in its prediction. The scoring rules presented below are all minimized by the narrowest possible distribution/interval containing the `true' parameter value $x$, but otherwise vary in the way in which they reward or punish qualities of the estimate.

\subsection{Top Hat Density}
Our first scoring rules will require no information other than an upper and lower bound. We introduce the top hat density weighting which is assumed uniform on the interval $\left[ x_1, x_2\right] = \left[ x_0 - \frac{1}{2}w,x_0 + \frac{1}{2}w\right]$:
$$
q(x) = 
\begin{cases}
0, & \text{if } x<x_1, \\
w^{-1}, & \text{if } x_1<x<x_2, \\
0, & \text{if } x>x_2.
\end{cases}
$$

\subsubsection{Log score}
We can treat the weight $q(x)$ as a probability distribution, and define the log score as 
\begin{equation}
 S_{\mbox{\scriptsize log }}(q|x)  := - \log\left(q(x)\right).
\end{equation}
Obviously, the log score will penalize you far too harshly if your interval does not contain the value $x$, i.e., it will return an infinite score. 

\subsubsection{Continuously ranked probability score (CRPS)}
The CRPS is a little more charitable:
$$
S_{CRPS}(q|x) = 
\begin{cases}
 -(x - x_0) - \frac{1}{6}w, & \text{if } x<x_1, \\
w^{-1}(x - x_0)^2 + \frac{1}{12}w, & \text{if } x_1<x<x_2, \\
x - x_0 - \frac{1}{6}w, & \text{if } x>x_2.
\end{cases}
$$ 
Once the interval contains $x$  and $w$ becomes large, the score goes like $w$. This means the penalty grows without bound.

\subsubsection{Brier score}
Another option is the Brier score which has a very simple form in this instance:
$$
S_B(q|x) = 
\begin{cases}
 w^{-1}, & \text{if } x<x_1,x>x_2 \\
-w^{-1}, & \text{if } x_1<x<x_2 \\
\end{cases}
$$
Note that this score is unbounded in the other direction: a wildly narrow interval will cost you if you are wrong.
\begin{figure}
\includegraphics[width=\columnwidth]{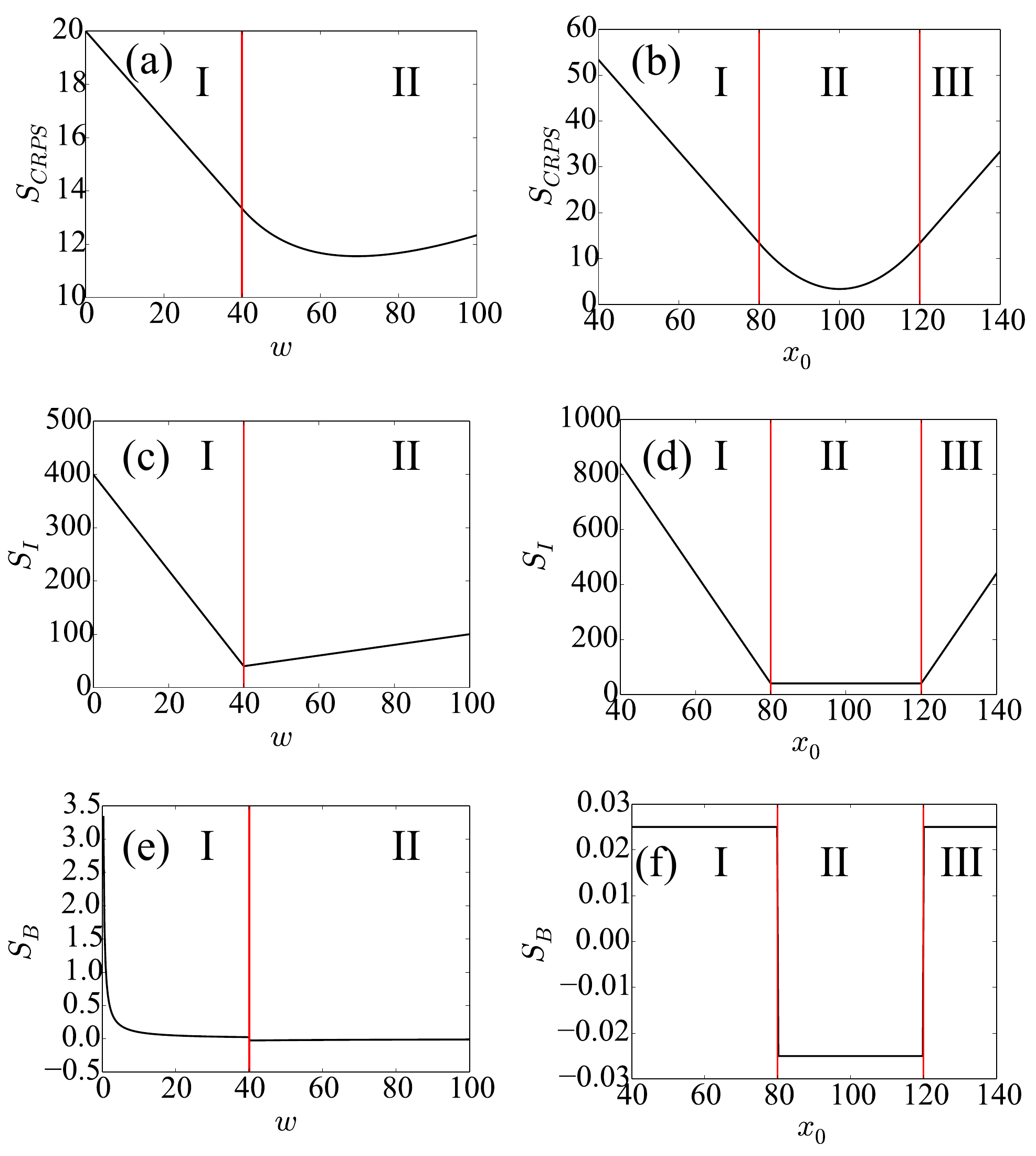}
\caption{(a), (c), (e): Scoring rules vs. interval widths $w$ for intervals $[x_1, x_2] = [x_0-w/2, x_0+w/2]$. Here, the interval centres are at $x_0=80$,  and the `true' value of $x=100$. (b), (d), (f): Scoring rules vs. interval centres, $x_0$, for fixed width $w=40$, where the `true' value $x=100$. Regions are labelled according to I: $x>x_2$, II: $x_1<x<x_2$ and III: $x<x_1$. The interval forecasts assume that the intervals represent 90\% prediction intervals.
} \label{Fig:srange}
\end{figure}

\subsection{Interval forecasts}
Instead of specifying a density, as in the top hat density, another approach is to issue an interval forecast. Suppose that the interval $\left[x_1, x_2\right]$ actually refers to the centered $(1 - \alpha) \times 100\%$ prediction interval. Then the following is a scoring rule:
$$
S_I(x_1,x_2|x) = 
\begin{cases}
 -\frac{2}{\alpha}(x - x_0) - (\frac{1}{\alpha}-1)w, & \text{if } x<x_1, \\
w, & \text{if } x_1<x<x_2, \\
 \frac{2}{\alpha}(x - x_0) - (\frac{1}{\alpha}-1)w, & \text{if } x>x_2.
\end{cases}
$$ 
Note that this is very similar to the CRPS - it is unbounded when $w$ is large - but that it depends on the precision $\alpha$. This precision dependence means that $S_I$ uses more information, which may allow for more informative scores (see Sec.\ \ref{Sec:Results}).

\begin{figure}
\includegraphics[width=\columnwidth]{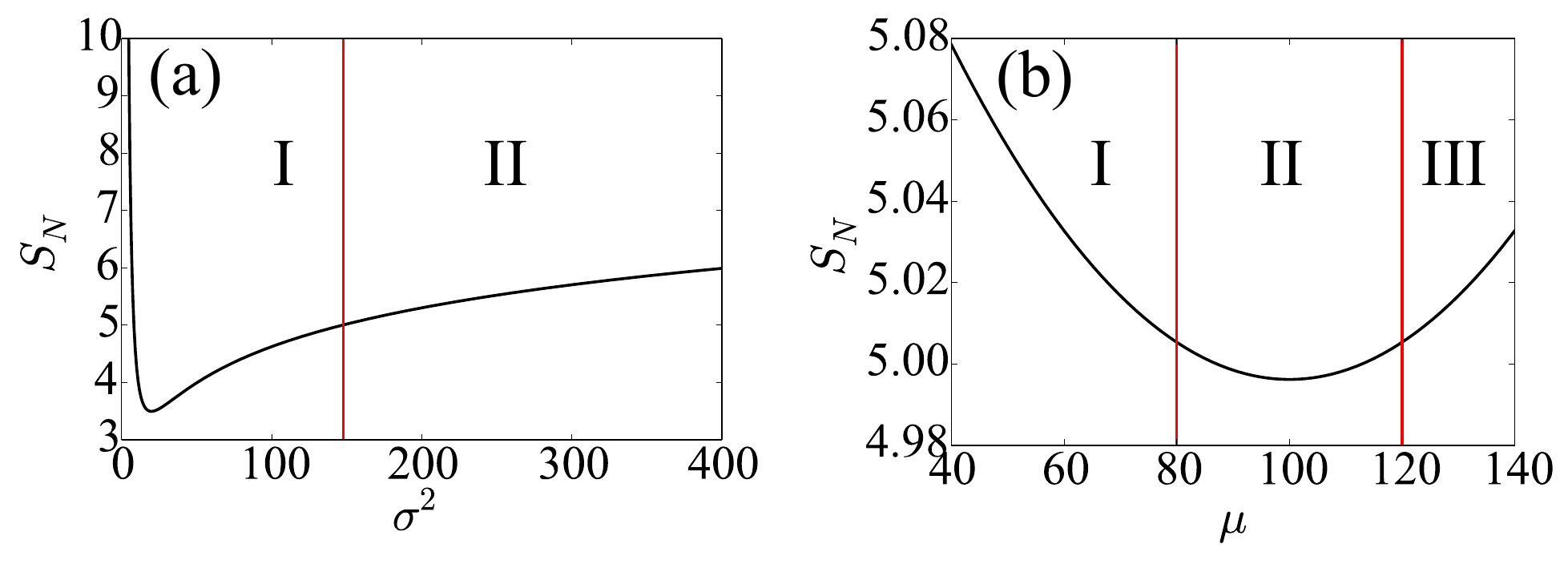}
\caption{(a) Norm score for a distribution of given variance, $\sigma^2$, centred around $\mu=80$, where the `true' value of $x=100$. (b) Norm score for a distribution of variance $\sigma^2\approx 148$ (corresponding to a Gaussian of 90\% prediction region of width 40) and mean = $x$. Regions are labelled according to I: $x>x_2$, II: $x_1<x<x_2$ and III: $x<x_1$, where $x_{1}$ and $x_2$ are the 5th and 95th percentiles of a Gaussian distribution of equivalent mean and variance.
} \label{Fig:srange_norm}
\end{figure}

\subsection{Moment Predictions}
There are some scoring rules that can be applied when only the mean, $\mu$, and variance, $\sigma^2$, are given. The
commonest one is essentially the log score applied to a normal density of the same mean and variance:
\begin{equation}
 S_N(\mu,\sigma|x) = \frac{1}{2}\left(\frac{x-\mu}{\sigma}\right)^2 + \log(\sigma) .
\end{equation} 
There is a technical caveat: this is not a strictly proper scoring rule. A plausibility argument for this scoring rule comes from noting that the normal density is the maximally entropic distribution with a given mean and variance \cite{Dawid:1999}.

\begin{figure}
\includegraphics[width=\columnwidth]{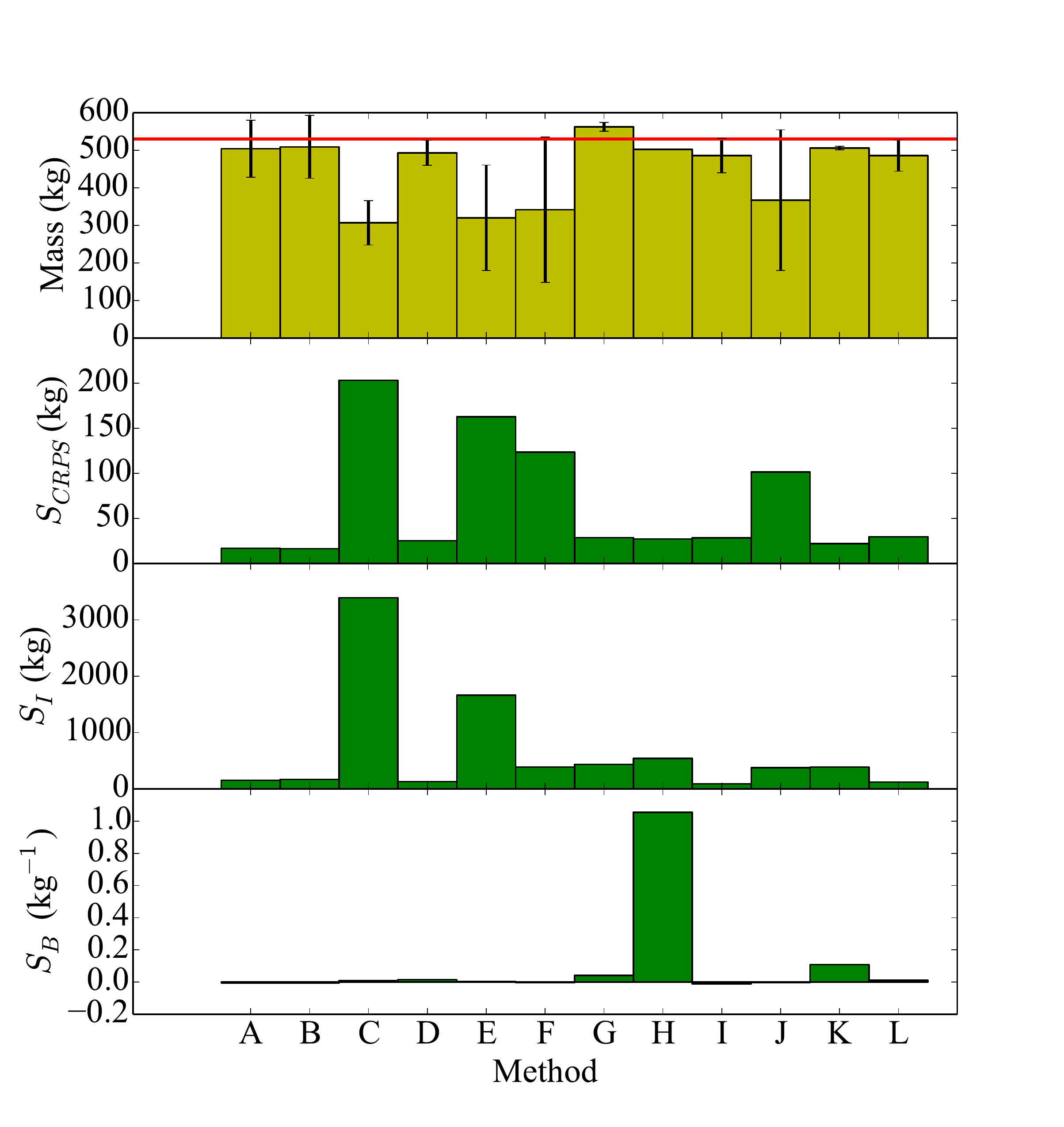}
\caption{Mass predictions (bars, top panel) from methods A-L (errorbars indicate the $90\%$ prediction intervals); the `true' mass is indicated by the horizontal line. Scores from different scoring rules (bottom three panels), for the corresponding predictions from methods A-L.
} \label{Fig:bars}
\end{figure}

\subsection{General remarks}
Figs.\ \ref{Fig:srange} and \ref{Fig:srange_norm} illustrate the difference in behaviour of the different scoring functions. 
All functions have similarities in their general features, such that they punish narrow intervals/small variances when the distribution centre $x_0$ (or $\mu$) is far from the `true' value $x$ (see the regions marked I in Figs.\ \ref{Fig:srange}(a), (c), (e) and Fig.\ \ref{Fig:srange_norm}(a)). However, $S_I$, $S_{CPRS}$ converge to a constant when $w\rightarrow 0$, and $x_0\neq x$, whereas $S_B$, $S_N$ diverge to infinity. The scoring rules also punish measurements with large widths when the distribution centre is close to the `true' value (see the regions marked II in the same figures): all scores increase as $w \rightarrow \infty$ (or $\sigma^2 \rightarrow \infty$); however, $S_i$, $S_{CPRS}$, $S_N$ diverge to infinity (albeit $S_N$ diverges rather slowly), whereas $S_B$ converges to 0 from below.

All rules punish measurements with $x_0$ (or $\mu$) far from the `true' value $x_0$ (see Figs.\ \ref{Fig:srange}(b), (d), (f) and Fig.\ \ref{Fig:srange_norm}(b)); however, $S_{I}$ is piecewise-linear in $x_0$ when $x$ is both outside and inside the prediction interval, $S_{CRPS}$ is piecewise quadratic inside and linear outside, $S_N$ is quadratic everywhere, and $S_B$ is piecewise constant - once $x$ is no longer contained in the prediction interval it makes no difference how far away it is.

\section{Experiment} \label{Sec:Experiment}
We apply our scoring rules to experimental data from a walkover weighing experiment for cattle. The forces recorded by load cells at the front and the back of a weighbridge were recorded during walkovers of 47 cows. The cows were also each weighed using a `static measurement' obtained when the cow was stopped on the weighbridge and sufficient time allowed for the cow to become still. Fourteen inference methods, labelled A-N produced 90\% prediction intervals for the mass given the force measurements. The static measurements were  assumed to give accurate measurements with uncertainties much smaller than those of the walkover measurements. This enabled the mass predictions from methods A-N to be scored against the static measurements. Details of the inference methods are outside the scope of this paper, but we describe the method labelled E below as a pedagogical example.

\subsection{Example method}
Method E is a simple-minded method of estimating a prediction interval from the timeseries of force data composing a walkover measurement. The method consists in taking the mean, $\mu_F$ and standard deviation, $\sigma_F$ of all forces $F$ for which $F/g>50$ kg. The confidence interval estimate is given as $\left[ \mu_F - \sigma_F,  \mu_F + \sigma_F \right]$. This method is expected to score badly for two reasons: firstly, the interval will be centred lower than the true mass due to the contributions of smaller forces during the time that the cow is stepping on and off the weighbridge; secondly, the interval width of $2\sigma_F$ is not expected to provide an optimal 90\% prediction region.

\section{Results and Discussion} \label{Sec:Results}
\subsection{Example walkover}
Figure \ref{Fig:bars} shows the predictions A-L, along with the scores $S_{CRPS}$, $S_{I}$ and $S_{B}$ for a single walkover in the experiment (note that, as expected, method E underestimates the mass; also, methods M and N have been omitted in this example).
Clearly the different scores punish different predictions. The CRPS score punishes the predictions from methods C, E, F and J most harshly - these produced the prediction intervals with centres furthest from the `true' mass. Notice that methods F and J, whose prediction intervals overlap the `true' mass are punished more harshly than, e.g., methods H and K whose intervals do not. Conversely, the Brier score punishes methods H and K most harshly, giving no credit for the proximity of the interval centre to the `true' value. The interval forecast appears to give more of a balance between the reward for accuracy of the interval centre and punishment for spurious precision indicated by the interval width. This improved balance between is due to specification of the precision $\alpha$ for the interval forecast allowing in some sense `the correct' weighting between the interval centre and width.
\subsection{Overall analysis}
\begin{figure}
\includegraphics[width=\columnwidth]{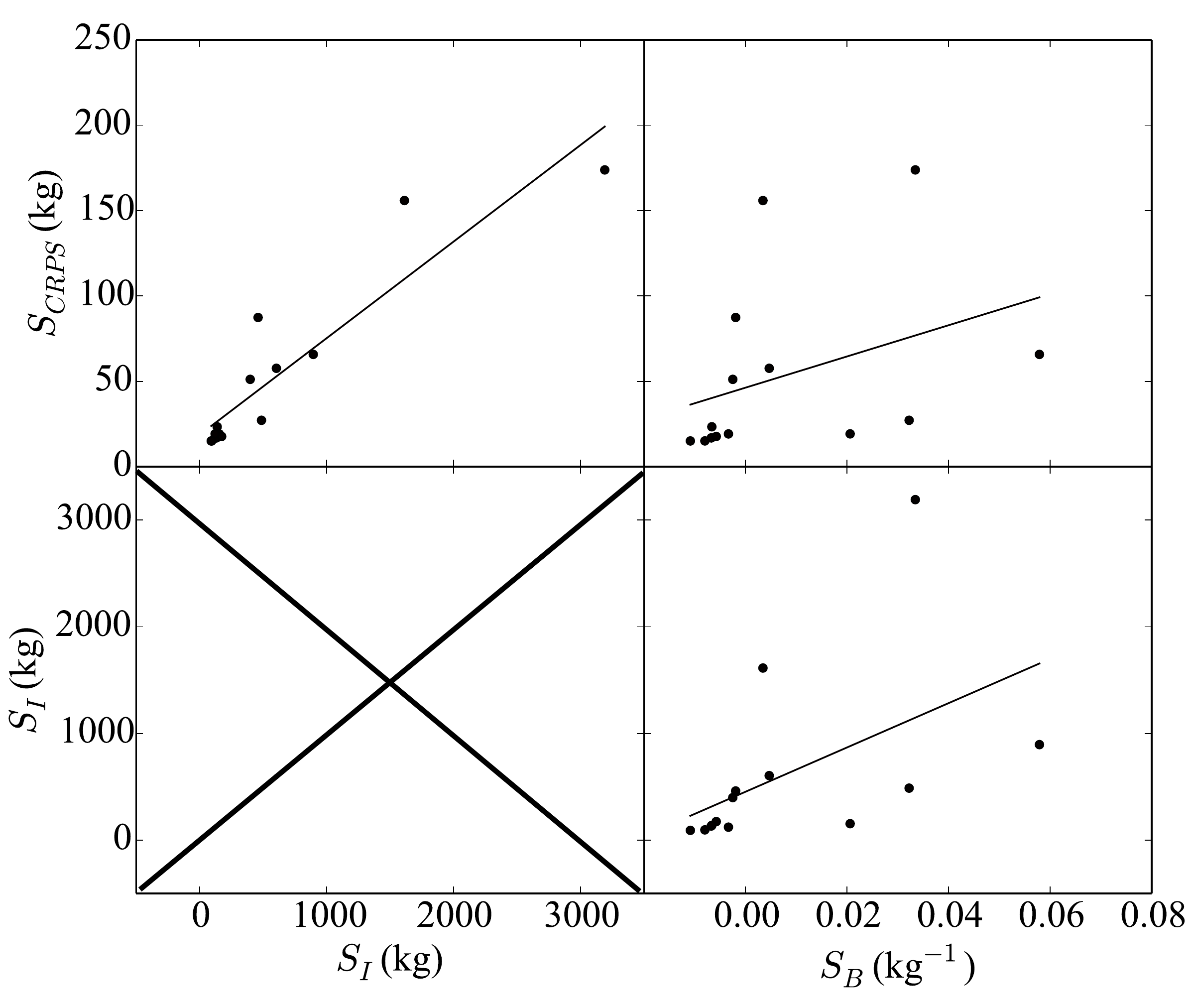}
\caption{Scatters of mean scores over the experimental walkover measurements by methods A-N. Lines represent best-fit by least-squares linear regression.
} \label{Fig:corro}
\end{figure}

\begin{table}[bpht!]
\caption{Methods and scores, $S$, corresponding to the mean scores over all experimental runs. Methods are ranked by $S_{CPRS}$. }
\label{Table:ranks}
\centering
\begin{tabular}{  l | l | l | l | l  }
\hline\hline
Rank & Method  & $S_{CRPS}$ (kg)  & $S_I$ (kg) & $S_B$ (kg$^{-1}$) \\ \hline 
1 & I & 15.1 & 91.4 & -0.0109 \\ 
2 & L & 15.1 & 95.8 & -0.00801 \\ 
3 & A & 16.9 & 132. & -0.00675 \\ 
4 & B & 17.7 & 174. & -0.00576 \\ 
5 & D & 19.2 & 121. & -0.00336 \\ 
6 & M & 19.2 & 154. & 0.0206 \\ 
7 & G & 23.4 & 138. & -0.00663 \\ 
8 & N & 27.2 & 487. & 0.0322 \\ 
9 & F & 51.2 & 398. & -0.00251 \\ 
10 & H & 57.6 & 604. & 0.00466 \\ 
11 & K & 65.8 & 895. & 0.0579 \\ 
12 & J & 87.4 & 461. & -0.00195 \\ 
13 & E & 156. & 1610 & 0.00343 \\ 
14 & C & 174. & 3190 & 0.0334 \\ 

\hline \hline
\end{tabular} 
\end{table}
Table \ref{Table:ranks} shows methods A-N with the median scores $S_{CRPS}$, $S_I$ and $S_B$ evaluated over all experimental walkovers for which the method successfully produced an output. The results can be used to rank the methods according to their scores. The choice of scoring rule affects the ranking order - although there is a clear correlation between the scores illustrated in Fig.\ \ref{Fig:corro}. Clearly, the strongest correlation is between $S_{CRPS}$ and $S_I$; the spread of in the scatter plot is due to the increased importance given to the uncertainty prediction by $S_I$ relative to $S_{CRPS}$. The Brier score is less well correlated with other scores, due to the the fundamental importance it places on the interval to contain the `true' value. However, there is still a clear correlation between the Brier score and the others. In unusual cases where methods either produce predictions close to the true value but report spuriously accurate precision, or produce prediction intervals with centres far from the true value but which overlap the true value, this correlation will become vanishingly small. Fortunately our method I is ranked number 1 by the $S_{CRPS}$, $S_I$ and $S_B$, indicating it as the superior method. Notice that our simple-minded method E scores badly in all measures.

\section{Conclusions} \label{Sec:Conclusions}
Ultimately, the choice of score used to rank sensor output should be motivated by preference of the user. If one places most importance on the accuracy of the measurement and least on the reported uncertainty, then the CRPS is the most useful scoring rule; conversely, if one is most concerned that the reported interval should contain the `true' value, then the Briar score is most useful. If one values both of these things, then the interval forecast represents the best candidate - this score makes use of the specified precision $\alpha$ to provide a more informative score. However, an ideal method will minimise all scores, and any method receiving a large value for any scoring rule should be regarded as problematic. As well as being useful for adjudicating between competing sensors, scoring rules may be useful for flagging problems in sensor calibration. For example, consistently high CRPS scores might indicate a systematic offset in the sensor output, whereas high Brier scores might indicate poor uncertainty evaluation.

\section*{Acknowledgment}
This work was funded by grant UOOX1208 from the Ministry of Business, Innovation \& Employment.


\bibliographystyle{IEEEtran}
\bibliography{IEEEabrv,./scoring_rules}
%
%
%

\end{document}